\documentclass[twocolumn,pra,showpacs,amsmath]{revtex4}

\usepackage{graphicx}

\begin{document}

\title
{Quantum Non-Demolition Detection of Polar Molecule Complexes:

Dimers, Trimers, Tetramers}

\author{Igor B. Mekhov}
\affiliation{University of Oxford, Department of Physics, Clarendon Laboratory, Parks Road, Oxford OX1 3PU, UK}

\begin{abstract}

The optical nondestructive method for in situ detection of the bound states of ultracold polar molecules is developed. It promises a minimally destructive measurement scheme up to a physically exciting quantum non-demolition (QND) level. The detection of molecular complexes beyond simple pairs of quantum particles (dimers, known, e.g., from the BEC-BCS theory) is suggested, including three-body (trimers) and four-body (tertramers) complexes trapped by one-dimensional tubes. The intensity of scattered light is sensitive to the molecule number fluctuations beyond the mean-density approximation. Such fluctuations are very different for various complexes, which leads to radically different light scattering. This type of research extends "quantum optics of quantum gases" to the field of ultracold molecules. Merging the quantum optical and ultracold gas problems will advance the experimental efforts towards the study of the light-matter interaction at its ultimate quantum level, where the quantizations of both light and matter are equally important.

\end{abstract}


\maketitle

\section{Introduction}

The study of ultracold polar molecules attracts significant attention because of their long-range anisotropic interaction, which can lead to the creation of exotic quantum phases of ultracold particles. The phase diagram is expected to be much richer than that for atomic gases with the short-range interaction. Recently, the existence of several few-body bound states of polar molecules has been proved for a low dimensional geometry \cite{PRL2011,arXiv2011}. Being important in the context of few-body physics, those results can help to get insight into the many-body problems as well \cite{Zoller}, where the elementary few-body building blocks can play a crucial role. For example, going beyond the two-body complexes and predicting the existence of bound states consisting of more than two particles (as trimers and tetramers), those results can modify the standard description of the BCS-BEC crossover in certain systems, which is usually based on the picture of pairs (i.e., the dimers) only. In contrast to extensively studied Efimov-type states \cite{R21} with short-range contact interaction, the states appearing due to the anisotropic long-range dipole-dipole interaction are less investigated.

\begin{figure}
\scalebox{0.39}[0.39]{\includegraphics{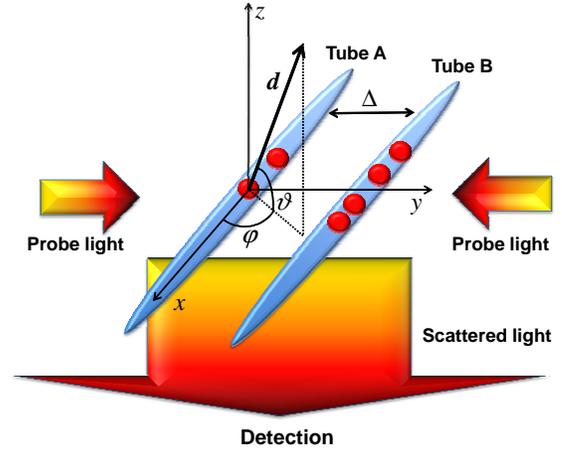}}
\caption{Setup. The molecules with dipole moment ${\bf d}$ are trapped in the potential of two 1D tubes. The probe and detection are in the plane perpendicular to the tubes.}
\end{figure}

The use of optical methods to detect the states of polar molecules promises the development of non-destructive in situ measurement schemes, which can be used to probe the system dynamics in real time. Moreover, as has been suggested in Refs. \cite{PRL2011,arXiv2011}, the optical non-destructive detection of ultracold molecules can be developed up to the physically exciting quantum non-demolition (QND) level. Such an ultimately quantum measurement scheme affects the quantum state in a minimally destructive way and triggers the intriguing fundamental questions about the quantum measurement back-action and the entanglement between the light and ultracold molecules \cite{PRA2009,PRL2009,LasPhys2010,LasPhys2011}. Other probing methods such as time-of-flight measurements or lattice shaking \cite{R22} are usually destructive. This paper provides further details about the QND measurements in ultracold molecular gases \cite{PRL2011,arXiv2011}. Focusing on a simple physical picture of the light-matter interaction, we show how the main characteristics of the light scattering can be estimated analytically, using a simple statistical approach. Moreover, those results should be valid even in the many-body systems with a large number of ultracold molecules, at least, in the low-density regime. This type of research extends the field of "quantum optics of quantum gases" \cite{NaturePhys,PRA2009} for the molecular species. Merging the quantum optical and ultracold gas problems will advance the experimental efforts \cite{Exp1,Exp2,Exp3,Exp4,EsslingerDicke,Kuhr,Ketterle2011} towards the study of the light-matter interaction at its ultimate quantum level.

\section{Light scattering from ultracold molecules in 1D}

We consider ultracold dipolar molecules trapped in the potential of two one-dimensional (1D) tubes (cf. Fig. 1). As described in details in Refs. \cite{PRL2011,arXiv2011}, even for the repulsive interaction between the molecules within each tube, they can form bound complexes due to the attractive dipole-dipole interaction with the molecules in a different tube. Thus, several repulsing molecules in one tube can be bound by the presence of a molecule in another tube, with which they interact attractively. The association of molecules into various stable complexes was proved: dimers "1-1" (with one molecule in each tube), trimers "1-2" (with one molecule in one tube and two molecules in the other tube) and tetramers "1-3" (with one molecule in one tube and three molecules in the other tube) and "2-2" (with two molecules in each tube).

The few-body complexes can be detected using light scattering. Recently, several nondestructive (in the sense of the quantum non-demolition, QND) schemes for measuring the properties of the many-body states in ultracold gases observing scattered light have been proposed \cite{PRL2007,PRA2007,LasPhys2009,PRA2009,Cirac08,Demler09,Polzik09}. Among them, the method developed in Refs.~\cite{PRL2007,PRA2007,LasPhys2009,PRA2009} is the most relevant to the present system, as it explicitly uses the sensitivity of light scattering to the relative position of the particles forming a complex. This is due to the constructive or destructive interference of the light waves scattered from the different particles. This method can be directly applied for extended periodic structures (many equidistantly spaced tubes or layers) and many-body systems, which makes the experimental realization promising. In contrast to Refs. \cite{Cirac08,Demler09,Polzik09} and experiments \cite{KetterleSpin1,KetterleSpin2} with spin ensembles, the original proposal of Refs. \cite{PRL2007,PRA2007,LasPhys2009} does not rely on any state-selective (e.g., spin-selective) light scattering, but is sensitive to the particle position.

We consider the scattering of the probe light with the amplitude given by the Rabi frequency $\Omega_p=d_0E_p/\hbar$ ($E_p$ is the probe-light electric field amplitude and $d_0$ is the induced dipole moment), cf. Fig.~1. To increase the signal, the scattered light can be collected by a cavity, and the photons leaking from the cavity are then measured. Alternatively, the measurement of photons scattered can be made in a far-field region without the use of a cavity.

Using the approach of the second quantization for the molecule-field operator (as it was formulated for atoms in Refs.~\cite{PRL2007,PRA2007}), the amplitude of the scattered light (i.e., the annihilation operator of the scattered photon) is given by

\begin{eqnarray}\label{1.A}
a_s=C\int d{\bf r} \hat{\Psi}^\dag({\bf r})u_s^*({\bf r})u_p({\bf r})\hat{\Psi}({\bf r}),
\end{eqnarray}
where $\hat{\Psi}({\bf r})$ is the matter-field operator at the point ${\bf r}$. For the free space scattering, the value of $C$ corresponds to the Rayleigh scattering \cite{Scully}. Adding a cavity to the setup the scattering is increased and $C=-ig_s\Omega_p/(\Delta_a\kappa)$ with $\kappa$ being the cavity decay rate, $g_s$ is the molecule-light coupling constant, and $\Delta_a$ is the light detuning from the resonance, cf. Refs.~\cite{PRL2007,PRA2007,LasPhys2009,PRA2009}. In Eq.~(\ref{1.A}), $u_{p,s}({\bf r})$ are the mode functions of probe and scattered light, which contain the information about the propagation directions of probe and scattered light waves with respect to the tube direction. For the simplest case of two traveling light waves, the product of two mode functions takes the well-known form from classical light scattering theory: $u^*_s({\bf r})u_p({\bf r})=\exp{[i({\bf k}_p-{\bf k}_s){\bf r}}]$, where ${\bf k}_{p,s}$ are the probe and scattered light wave vectors.

One can express the matter-field operator in the basis of the functions corresponding to the transverse distribution of molecules within two tubes A and B:
\begin{eqnarray}\label{1.B}
\hat{\Psi}({\bf r})=\hat{\Psi}_A(x)w(\rho-\rho_A)+\hat{\Psi}_B(x)w(\rho-\rho_B),
\end{eqnarray}
where $\hat{\Psi}_{A,B}(x)$ are the matter-field operators within each tube with the coordinate $x$ alone the tube, where the molecules can move (cf. Fig. 1); $w(\rho)$ gives the distribution of a molecule in the transverse direction ($\rho$ is the transverse coordinate). Substituting this expression in Eq.~(\ref{1.A}), we can describe the light scattering taking into account the possible overlap of the molecules between two tubes (overlapping $w(\rho-\rho_A)$ and $w(\rho-\rho_B)$) and the nontrivial overlap between the molecule distribution $w(\rho)$ and the light modes $u_{p,s}({\bf r})$. However, following Refs. \cite{PRL2011,arXiv2011}, we assume that two tubes do not overlap at all, and they are well localized with respect to the light wave.

Thus, after several assumptions (the small tube radius, far off-resonant light scattering, detection in the far field zone), the light scattering has a simple physical interpretation. The scattered light amplitude is given by the sum of the light amplitudes, scattered from each molecule (cf. Fig.~1). Each term has a phase and amplitude coefficient depending on the position of the molecule as well as on the  direction and amplitude of the incoming and outgoing light waves:
\begin{eqnarray}\label{1.5}
a_s=C\sum_{i=A,B}\int dx \hat{n}_i(x) u^*_s(x,\rho_i)u_p(x,\rho_i),
\end{eqnarray}
where the sum is over two tubes A and B, $\hat{n}_i(x)=\hat{\Psi}^\dag_i(x)\hat{\Psi}_i(x)$ is the operator of particle linear density. In Eq.~(\ref{1.5}), $u_{p,s}(x,\rho_i)$ are the mode functions of probe and scattered light at the tube positions $\rho_{A,B}$.

Equation~(\ref{1.5}) is valid for any optical geometry and can describe the angular distribution of the scattered light. However, an important conclusion of Refs.~\cite{PRL2007,PRA2007,LasPhys2009,PRA2009} was that some information about the many-body state can be obtained even by a simple measurement of the photon number scattered at a single particular angle, which is fully enough for our purpose. Moreover, as it was shown, the particularly convenient angle of measurement corresponds to the direction of a diffraction minimum, rather than Bragg angle (diffraction maximum). At the directions of diffraction minimum any classical (possibly very strong) scattering is suppressed, and the light signal exclusively reflects the quantum fluctuations of the particles.

We now fix the optical geometry as follows (cf. Fig. 1). The incoming probe light is a traveling or standing wave propagating at the direction perpendicular to the tubes, which gives $u_{p}({\bf r})=R(x)\exp(ik_py)$ (for the traveling wave) or $u_{p}({\bf r})=R(x)\cos(k_py)$ (for the standing wave) and includes the transverse probe profile $R(x)$ of an effective width $W$. To perform the measurements at the direction of a diffraction minimum, the scattered light is measured along $z$ direction. For the free space detection, or the traveling-wave cavity, this gives $u_{s}({\bf r})=\exp(ik_sz)$, while for the case of a standing wave cavity, $u_{s}({\bf r})=\cos(k_sz)$. Without loss of generality, we can assume $u_{s}({\bf r})=1$ at the tube position $z=0$. The absolute values of the wave vectors are equal to their vacuum quantities $k_{p,s}=2\pi/\lambda_\text{light}$.

An important property of such a configuration (illumination and detection at the directions perpendicular to the tubes), is that all molecules within one tube scatter light with the same phase independently of their longitudinal position $x$ within the tube. Thus, the light scattered from the molecules within one tube interferes fully constructively. As a consequence, all molecules within two different tubes scatter light with a fixed phase difference with respect to each other. Due to this fact, the averaging over the probabilistic position of the complex does not involve the light phase and all complexes of the same type scatter light identically. Moreover, averaging over the probabilistic relative positions within each complex does not involve the dependence on the light phase as well. At other directions, both those kinds of phase averaging are important and would decrease the optical signal and the distinguishability of the complex types. The simple scattering picture also allows the generalization of the model for an array of several tubes.

The operator of the light amplitude reduces to
\begin{eqnarray}\label{1.6}
a_s=C\left(u_p(y_A)\hat{N}_A(W)+u_p(y_B)\hat{N}_B(W)\right),
\end{eqnarray}
where $\hat{N}_{A,B}(W)$ are the operators of the effective particle numbers in the tubes A and B within the region illuminated by the laser beam,
\begin{eqnarray}\label{1.7a}
\hat{N}_{A,B}(W)=\int_{-\infty}^{\infty}\hat{n}_{A,B}(x)R(x)dx.
\end{eqnarray}
If the laser profile can be approximated by a constant in the interval $(-W/2, W/2)$, the operators $\hat{N}_{A,B}(W)$ exactly correspond to the atom number operators in two tubes within the laser beam.

The classical condition of the diffraction minimum is fulfilled, when the expectation value of the light-amplitude operator (\ref{1.6}) is zero due to the perfect cancelation of the expectation values of two terms in Eq.~(\ref{1.6}) (i.e. the total destructive interference between the scatterers in two tubes). This is achieved for $u_p(y_B)/u_p(y_A)=-\langle\hat{N}_A\rangle/\langle\hat{N}_B\rangle$. We introduce the atom number ratio $\alpha=\langle\hat{N}_A\rangle/\langle\hat{N}_B\rangle$. For the equal mean atom numbers (the few-body complexes 1-1 and 2-2), the optical geometry should be chosen such that $u_s(y_B)/u_s(y_A)=-1$, which can be achieved if, e.g., the tube spacing is the half of the light wavelength, $\Delta=\lambda_\text{light}/2$. For the few-body complex 1-2, $\alpha=1/2$, and the diffraction minimum is achieved if the light wavelength and tube spacing satisfy the condition $\cos(k_py_B)/\cos(k_py_A)=-1/2$. This can be achieved, e.g., if the position of the tube A corresponds to the antinode of the standing wave $\cos(k_py_A)=1$, while that of tube B corresponds to $k_py_B=2\pi/3$ or $4\pi/3$, leading to the ratios between the tube spacing and light wavelength as $\Delta=\lambda_\text{light}/3$ or $2\lambda_\text{light}/3$. Similarly, for the 1-3 complex, that ratio can be $\Delta\approx 0.3\lambda_\text{light}$ or $0.7\lambda_\text{light}$. All those example ratios can be indeed larger, taking into account the periodicity of the light wave.

The expectation value of number of photons scattered at the direction of diffraction minimum $n_{\Phi}$ is then given by
\begin{eqnarray}\label{1.7}
n_{\Phi}=\langle a^\dag_s a_s\rangle = \nonumber \\
\left|C\right|^2 \left|u_p(y_A)\right|^2
\left\langle\left(\hat{N}_A(W)-\alpha\hat{N}_B(W)\right)^2\right\rangle,
\end{eqnarray}
where $u_p(y_A)$ can be easily chosen as 1. This expression manifests that the number of photons scattered in the diffraction minimum is proportional to the second moment of the "rated" particle number difference between two tubes in the laser-illuminated region. The mean light amplitude is sensitive to the mean values of the particle number and is precisely zero at the diffraction minimum: $\langle a_s\rangle \sim \left\langle\left(\hat{N}_A(W)-\alpha\hat{N}_B(W)\right)\right\rangle=0$. However, in general, the photon number (\ref{1.7}) is non-zero. It directly reflects the particle number fluctuations and correlations between the tubes. Thus, the number of photons reflects the quantum state of ultracold molecules.

\section{Applications for dimers, trimers and tetramers}

\begin{figure}
\scalebox{0.9}[0.9]{\includegraphics{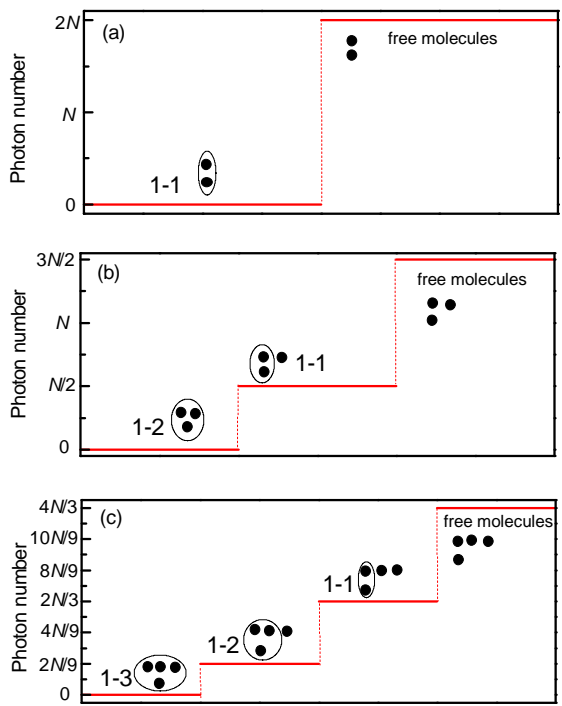}}
\caption{QND measurement of ultracold polar molecule complexes. Intensity of scattered light (i.e. the relative photon number $n_{\Phi}/\left|C\right|^2$) depending on the existence of various few-body complexes. The variable on the horizontal axis is schematic. It can correspond to several parameters, which allow to scan the system through the regimes, were different complexes exist (e.g., the dipole orientation angle or dipole-dipole interaction strength as shown in Refs. \cite{PRL2011,arXiv2011}). (a) Dissociation of dimers "1-1" into free molecules corresponds to the change of light intensity from $n_{\Phi}/\left|C\right|^2=0$ to $n_{\Phi}/\left|C\right|^2=2N$. (b) Dissociation of trimers "1-2" into dimers "1-1" and free molecules, and then into all free molecules corresponds to the intensity jumps as $n_{\Phi}/\left|C\right|^2=0$, $n_{\Phi}/\left|C\right|^2=N/2$ and $n_{\Phi}/\left|C\right|^2=3N/2$. (c) Dissociation of tetramers "1-3" into trimers "1-2" and free molecules, then into dimers "1-1" and free molecules, and finally into all free molecules correponds to the intensity values $n_{\Phi}/\left|C\right|^2=0$, $n_{\Phi}/\left|C\right|^2=2N/9$, $n_{\Phi}/\left|C\right|^2=6N/9$ and $n_{\Phi}/\left|C\right|^2=12N/9$. Inversely, the association of those complexes will correspond to the suppression of light scattered into the diffraction minimum.}
\end{figure}

In Ref. \cite{PRL2011,arXiv2011}, we presented the results of numerical simulations for light scattering from few-body complexes for particular parameters. We have shown that, while the photon number in the diffraction minimum is zero for a bound state, it immediately increases, when the complex dissociates into a smaller complex and a free molecule. Although after such a dissociation, the mean particle number stays the same (and the light amplitude would not change), the fluctuations of the particle number inside the laser beam change strongly after the dissociation: instead of one bound complex, one gets another complex and a free particle, whose positions are uncorrelated. The particle fluctuations increase the intensity of the scattered light.

In this paper, we demonstrate that the values of light intensity for stable complexes and free molecules can be estimated analytically using the statistical calculations. Such estimations are valid for many molecules in each tube (at least in the low-density regime) and agree well with the numerical simulations made for real systems, but the tiny number of molecules per tube \cite{PRL2011,arXiv2011}. The approach developed in this papers also gives a possibility to get a deeper physical insight into the problem. Although the development of modern trapping techniques targets the manipulation of ultracold atoms at a single-particle level \cite{R11}, the many-particle realization is still more realistic.

Expression (\ref{1.7}) can be written in the form
\begin{eqnarray}\label{1.N1}
n_{\Phi}/\left|C\right|^2 = \langle(\hat{N}_A -\alpha\hat{N}_B)^2\rangle = \nonumber \\ \langle\hat{N}^2_A\rangle +\langle\hat{N}^2_B\rangle - 2\alpha \langle\hat{N}_A\hat{N}_B\rangle,
\end{eqnarray}
which underlines the correlations between the molecule numbers in two different tubes.

Let us start with the example of dimers "1-1" and consider the equal number of molecules in two tubes ($\langle\hat{N}_A\rangle=\langle\hat{N}_B\rangle=N$, $\alpha=1$). When all molecules are strongly bound into dimers, they appear within the laser beam only in pairs, or do not appear there at all. Thus, the fluctuations of the molecule number difference is zero (one can think about the two number operators as identical ones, $\hat{N}_A=\hat{N}_B$, i.e., all their moments coincide) and so does the light intensity: $n_{\Phi}/\left|C\right|^2 = \langle(\hat{N}_A -\alpha\hat{N}_B)^2\rangle = \langle(\hat{N}_A -\alpha\hat{N}_B)\rangle=0$. On the other hand, when a dimer dissociates into two independent free molecules, the two operators are different, and the term with the intertube correlation function in Eq. (\ref{1.N1}) decorrelates into a product: $\langle\hat{N}_A\hat{N}_B\rangle=\langle\hat{N}_A\rangle\langle\hat{N}_B\rangle=N^2$. One can assume that the number fluctuations of the independent free molecules are Poissonian, $\langle\hat{N}^2_{A,B}\rangle=\langle\hat{N}_{A,B}\rangle^2+\langle\hat{N}_{A,B}\rangle=N^2+N$. Then, the number of scattered photons Eq. (\ref{1.N1}) gets $n_{\Phi}/\left|C\right|^2 = 2N$.

Therefore, we see that the light intensity jumps from zero to $n_{\Phi}/\left|C\right|^2 = 2N$, when the dimers dissociate into free molecules. Such a change of light intensity for two different phases of ultracold molecules is schematically demonstrated in Fig. 2(a). Physically, the strongly bound complex does not scatter light, because the geometry corresponds to the diffraction minimum. Thus, the fluctuation of the complex number within the laser beam does not change the light intensity (it is zero if both the complex is within the beam, and, obviously, outside the beam). However, when the complex dissociate into two independent species (two free molecules in this example), the species can be within or outside the beam independently from each other. Thus, the condition of the total diffraction minimum is not satisfied anymore, because, probabilistically, the numbers of molecules within the beam can be nonequal in two tubes (even though they are always equal in average) and the complete destructive interference of light is not possible anymore. Note, that this result agrees very well with the numerical calculations presented in Ref. \cite{arXiv2011} carried out for two molecules in two tubes. Those numerical results indeed show not only the constant values of the light intensity, but also describe the continuous transition between them, when the dimer dissociates.

Let us now consider the case of trimers "1-2", when the populations of two tubes are imbalanced: $\langle\hat{N}_A\rangle=N$, $\langle\hat{N}_B\rangle=2N$, $\alpha=1/2$. When the molecules are strongly bound into a trimers, they appear in the laser beam only all three together, or do not appear at all (Here we indeed neglect the small effects when the trimer is large and can overlap with the laser beam only partially. This however could be captured by the numerical simulations in Ref. \cite{PRL2011,arXiv2011}, and was shown to introduce only small corrections to the result.) Therefore, the fluctuations of the operator $(\hat{N}_A -\alpha\hat{N}_B)^2$ are zero and the number of scattered photons is zero as well: $n_{\Phi}/\left|C\right|^2 = \langle(\hat{N}_A -1/2\hat{N}_B)^2\rangle = \langle(\hat{N}_A -1/2\hat{N}_B)\rangle=0$.

The trimer can dissociate into a dimer "1-1" and a free particle, which are independent from each other. The operator of the number of particles in the tube B can be split into two parts: $\hat{N}_B=\hat{N}^D_B+\hat{N}^F_B$, where the operator $\hat{N}_B=\hat{N}^D_B$ corresponds to the molecules, which form a dimer with another molecule in the tube A, and $\hat{N}_B=\hat{N}^F_B$ corresponds to the free molecules. To calculate the expectation value for the photon number, we can group the molecule number operators in Eq. (\ref{1.N1}) such that they would correspond to the same species (dimers or free molecules). Then, $\hat{N}_A-1/2\hat{N}_B=\hat{N}_A-1/2\hat{N}^D_B-1/2\hat{N}^F_B=1/2(\hat{N}^D- \hat{N}^F)$, where we introduced the operators for the number of dimers, $\hat{N}^D=\hat{N}^D_B=\hat{N}_A$, and number of free molecules, $\hat{N}^F=\hat{N}^F_B$.

After introducing the operators for different independent species (dimers and free molecules), we can calculate the expectation value in Eq. (\ref{1.N1}), assuming that the species are uncorrelated ($\langle\hat{N}^D\hat{N}^F\rangle=\langle\hat{N}^D\rangle\langle\hat{N}^F\rangle=N^2$) and each of them displays the Poissonian fluctuations ($\langle(\hat{N}^{D,F})^2\rangle=\langle\hat{N}^{D,F}\rangle^2+\langle\hat{N}^{D,F}\rangle= N^2+N$). The result reads: $n_{\Phi}/\left|C\right|^2 = N/2$. So, we see, how the light intensity jumps from zero to this non-zero value, when the trimer dissociates into a dimer and a free molecule.

Those dimers and free molecules can dissociate further into three independent molecules. Taking into account the mean values of the free molecules in two tubes, $\langle\hat{N}^{F}_A\rangle=N$ and  $\langle\hat{N}^{F}_B\rangle=2N$, the expectation value of the light intensity reads $n_{\Phi}/\left|C\right|^2 = 3N/2$. That is, it jumps further upwards.

The consecutive dissociation of the trimers is schematically shown in Fig. 2(b). All three phases can be distinguished by the light intensity: it is zero for bound trimers, proportional to $N/2$ for dimers and free particles, and to $3N/2$ for all free particles. This result agrees with the numerical simulations \cite{PRL2011,arXiv2011}.

Let us now expand the consideration for the the case of tetramers "1-3", when the populations of two tubes are imbalanced: $\langle\hat{N}_A\rangle=N$, $\langle\hat{N}_B\rangle=3N$, $\alpha=1/3$. The numerical simulations for that situation were not reported in Refs. \cite{PRL2011,arXiv2011}. As before, when the complex is strongly bound, it does not scatter light into the diffraction minimum and $n_{\Phi}/\left|C\right|^2 = \langle(\hat{N}_A -1/3\hat{N}_B)^2\rangle = 0$. The following steps of a tetramer dissociation are possible: 1) a trimer "1-2" and a free molecule, 2) a dimer "1-1" and two free molecules and 3) three free molecules.

The tetramers first dissociate into the trimers "1-2" and free molecules. Proceeding as before, the number operator in the tube B can be split into two statistically independent operators: $\hat{N}_B=\hat{N}^T_B+\hat{N}^F_B$, where $\hat{N}^T_B$ corresponds to the molecules in the tube B, which form trimers with molecules in A, and $\hat{N}^F_B$ corresponds to free molecules. As before, we introduces the operator of the trimer number $\hat{N}^T$. All molecules in the tube A participate in the trimer creation: $\hat{N}_A=\hat{N}^T$, while the number of molecules forming the trimer in the tube B is two times larger: $\hat{N}^T_B=2\hat{N}^T$. Proceeding as before, assuming that the trimers and free molecules are not correlated ($\langle\hat{N}^T\hat{N}^F\rangle=\langle\hat{N}^T\rangle\langle\hat{N}^F\rangle=N^2$), and obey the Poissonian fluctuations, we arrive to the photon number as $n_{\Phi}/\left|C\right|^2 = \langle(\hat{N}_A -1/3\hat{N}_B)^2\rangle = 2N/9$.

After that, this four-body state can dissociate further into dimers "1-1" and two free molecules. All molecules in the tube A will form the dimer, $\hat{N}_A=\hat{N}^D$, and the number of molecules from the tube B forming the dimers will be the same, $\hat{N}^D_B=\hat{N}^D$. In this case, the mean values are: $\langle\hat{N}^D\rangle=N$, $\langle\hat{N}^F\rangle=2N$. The photon number jumps upwards: $n_{\Phi}/\left|C\right|^2 = \langle(\hat{N}_A -1/3\hat{N}_B)^2\rangle = 2N/3$.

Similarly, the last step of dissociation leading to all free molecules increases the intensity of scattered light further due to even stronger fluctuations of the molecule number within the beam: $n_{\Phi}/\left|C\right|^2 = 4N/3$. (To derive this expression, note that $\langle\hat{N}^F_A\rangle=N$, while $\langle\hat{N}^F_B\rangle=3N$).

The dependence of the light intensity on the molecule state is schematically shown in Fig. 2(c). The plateaus with four different values are expected: $n_{\Phi}/\left|C\right|^2=0$ for the tetramers "1-3", $n_{\Phi}/\left|C\right|^2=2N/9$ for trimer "1-2" and free molecules, $n_{\Phi}/\left|C\right|^2=6N/9$ for dimers "1-1" and free molecules, and $n_{\Phi}/\left|C\right|^2=12N/9$ for the totally dissociated system.

We have seen that the dissociation of a complex increases the particle number fluctuations, which leads to the jump of the light intensity. Inversely, the observation of the consecutive association would correspond to the stepwise suppression of the light intensity, which reflects the decrease of the number fluctuations. As shown in Ref. \cite{PRL2011,arXiv2011}, to go through all those stages one can change the orientation of the dipoles (e.g., the angle $\theta$ in Fig. 1), or the strength of the dipole-dipole interaction between the molecules. Interestingly, in contrast to the light intensity in the diffraction minimum, the mean light amplitude would not change at all and would stay zero for all states considered above. This is an example of a quantum optical problem, where one has a zero light amplitude $\langle a_s\rangle=0$, but non-zero photon number $\langle a^\dag_s a_s\rangle \ne |\langle a_s\rangle|^2$ due to the matter-induced photon fluctuations.

\section{Conclusions}

The optical nondestructive scheme for probing bound states of ultracold polar molecules is presented. Based on the off-resonant light scattering it promises the in situ measurement of the molecular dynamics in real time up to a physically exciting QND level. The detection of association and dissociation of molecular pairs (dimers), three-body states (trimers) and four-body states (tetramers) has been demonstrated. In contrast to other QND schemes \cite{Cirac08,Demler09,Polzik09,KetterleSpin1,KetterleSpin2} requiring the state-selective (e.g. spin-selective) light scattering, this method is originally based on the proposal of Refs. \cite{PRL2007,PRA2007,LasPhys2009} and is not sensitive to the internal-level structure, which is its advantage. The light scattering directly reflects the relative spatial positions of the complex parts and measures the quantum fluctuations of the molecule numbers beyond the mean-density approximation. Development of such QND techniques opens the field of "quantum optics of quantum gases" \cite{NaturePhys,PRA2009} for ultracold molecular gases and raises intriguing questions about the quantum measurement back-action and preparation of the exotic many-body phases using the entanglement between the light and many-body molecular states \cite{EPJD2008,PRL2009,PRA2009,LasPhys2010,LasPhys2011}. Merging the quantum optical and ultracold gas problems will advance the experimental efforts \cite{Exp1,Exp2,Exp3,Exp4,EsslingerDicke,Kuhr,Ketterle2011} towards the study of the light-matter interaction at its ultimate quantum level, where the quantum natures of both light and matter play equally important roles.

\section*{Acknowledgement}

The stimulating discussions with B. Wunsch, N. Zinner and E. Demler are appreciated. This work was supported by the EPSRC project EP/I004394/1.

\newpage

\end{document}